\documentclass[]{article}

\usepackage[utf8]{inputenc}
\usepackage{natbib}
\usepackage{amssymb}
\usepackage{amsmath}
\usepackage{epsfig}
\usepackage{url}

\title{Error Rates for Unvalidated Medical Age Assessment Procedures}
\author{Petter Mostad\footnote{Mathematical Sciences, Chalmers University of Technology and Gothenburg University, Sweden 
(mostad@chalmers.se)} \and Fredrik Tamsen\footnote{
Forensic Medicine, Department of Surgical Sciences, Uppsala University, Sweden}
}

\date{\today}

\begin{document}

\maketitle

\begin{abstract}
During 2014-15 Sweden received asylum applications from more than 240.000 people, of which more than 40.000 were termed
unaccompanied minors. In a large number of cases, claims by asylum seekers of 
being below 18 years were not trusted by Swedish authorities. To handle the situation, 
the Swedish national board of forensic medicine (Rätts\-medicinalverket, RMV) was assigned by the government to create a centralized system for medical age assessments.
RMV introduced a procedure including two biological age indicators; x-ray of the third molars and magnetic resonance imaging of the distal femoral epiphysis. In 2017 a total of 9617 males and 337 females were subjected to this procedure. No 
validation study for the procedure was however published, and the observed number of cases with different maturity combinations in teeth and femur were unexpected given the claims originally made by RMV. Such unexpected results might be caused by systematic errors and need to be analyzed thoroughly.

In the present paper we present a general stochastic model enabling us to study which combinations of age indicator 
model parameters and age population profiles are consistent with the observed 2017 data for males. We find that, 
contrary to some RMV claims, maturity of the femur, as observed by RMV, appears on average well before maturity of teeth. 
Although results naturally contain much uncertainty, 
we find that classification error rates for certain groups who based on the RMV procedure are 
classified as above 18 years may be around 10-30\%, possibly as high as 50\%.

\end{abstract}

\section{Introduction}
\label{secIntro}

In medical age assessments certain biological processes that develop during childhood in a predictable sequence are used to assess a person's chronological age. Examples of such processes are the development of teeth, bones, and sexual maturity.
We will use in this paper the term "age indicator" to mean 
any observed biological feature that develops through a series of clearly defined states over an age period 
relevant for age assessment.

Medical age assessments have been used for a long time in many countries, but are always associated with debate. This debate can be divided into two main lines. One concerns the very use of medical age assessment. Critics argue that since biological processes always show such a wide variation in populations, biological states can never be used to make sufficiently certain assessments of chronological age. Others might argue that it is the biological and not the chronological age that is the most relevant for the needs of an individual. 

In most countries the age is important for the rule of law. Many countries view the age of 18 as being the border between childhood and adulthood. Children have other needs and rights than do adults, and punishments for crimes might differ whether the perpetrator is below or above 18. In the case of asylum seekers, children are to be treated differently according to international conventions. One might therefore argue that if a person's age is unknown, a medical age assessment may be necessary in order to protect the privileges of children.

The other line of debate concerns which methods are appropriate to use. A compilation of methods used in the European Union shows that most countries use two or more age indicators \cite{EASO2018}. There are variations between countries, but the two most commonly used methods are dental x-ray and x-ray imaging of the hand/wrist. Another commonly used indicator is x-ray imaging of the collar bone. 
These three age indicators are all included in the recommendations by The Study Group on Forensic Age Diagnostics (Arbeitsgemeinschaft für Forensische Altersdiagnostik; AGFAD), an international assembly of experts that has worked on this issue for the last 18 years \cite{schmeling2008criteria}, \cite{schmeling2016forensic}. 

During the last few years, many asylum seekers in Sweden have claimed to be under 18 but have not been able to convince
authorities about their claim. To be able to treat children as children, and to not give child privileges to adults, the government of Sweden has decided to offer the possibility to make a medical age assessment in these cases. When 
a wave of asylum seekers arrived in 2014-15, there was no generally accepted system for medical ages assessments in Sweden, and Rättsmedicinalverket (RMV) was assigned by the government to create one.

The method\footnote{https://www.rmv.se/verksamheter/medicinska-aldersbedomningar/metoder/} chosen by RMV uses two age indicators: Magnetic resonance imaging of the distal femur (MRI knee), and x-ray imaging of the third molars in the mandible (x-ray teeth) 
\cite{RMV2017}. 
The MRI knee and x-ray teeth are independently evaluated by two radiologists and two dentists, 
respectively. For the knee to be assessed as mature, both radiologists must agree on this assessment. 
If they disagree, the knee is assessed as immature. The same procedure is used for teeth. 
These assessments are then combined in such a way that if either the knee or the teeth are mature, 
the individual is assessed as being 18 years or older\footnote{The exact wording of the conclusions produced by RMV have several forms. However, as these conclusions are then mapped by the Swedish migratory authority to decisions about 
age, the exact wording is of little consequence. We will in this paper simply refer to assessment of above or below 18 years.}. 
During 2017 a total of 9617 males and 337 females were subjected to this age assessment procedure. 
The results for males are given in Table~\ref{tab:data}. 
In 2018 RMV changed their assessments for females, since a new study showed that the majority of females aged 16 and 17 years had mature knees, see \cite{ottow2017forensic}, \cite{tamsen2017}. Females now need mature teeth to be assessed as being 18 years or older. In this paper, we will only study the RMV data for males. 

\begin{table}
\begin{center}
\begin{tabular}{|l|c|c|c|c|} \hline
 & Knees mature & Knees immature & No data knees & SUM \\ \hline
Teeth mature     & 4176 & 348 & 187   & 4711 \\ \hline
Teeth immature & 1735 & 1087 & 83   & 2905 \\ \hline
No data teeth                        & 1364 & 237  & 63    & 1664 \\ \hline
SUM                            & 7275 & 1672 & 333 & 9280    \\ \hline
\end{tabular}
\caption{Results for the 9280 males submitted to the RMV procedure during 2017.}
\label{tab:data}
\end{center}
\end{table}

The maturity of the teeth is assessed according to the stages of Demirjian, in which a tooth can be in one of eight stages A-H, see \cite{demirjian1973new}. H is the final stage and the teeth are 
termed 'mature' if at least one of the mandibular third molars are assessed as being in this stage. 
The knee is assessed as 'mature' if it has reached stage 4 or 5 according to the classification by Schmeling \cite{kramer2014forensic}. 
The different stages of immature teeth and knees are not used in the age assessments made by RMV. So, if the knee is mature, it doesn’t matter if the most developed examined tooth is in stage G (one stage from mature) or stage F (two stages from mature), the age assessment is still the same. 

%[Här vore det kanske bra att infoga en figur som visar de möjliga kombinationerna i RMVs grön-röda färger och tillhörande resultat. Jag kan fila på en sådan om du tycker det är en bra idé. Jag tycker det är en bra idé. Fast jag undrar på om vi skall försöka undvika färger i figurer.]

To date there exist six original articles and one letter to the editor on age assessment with MRI knee: 
\cite{dedouit2012age}, \cite{kramer2014forensic}, \cite{saint2015contribution}, \cite{ekizoglu2016forensic}, 
\cite{fan2016forensic}, \cite{ottow2017forensic}, \cite{vieth2018forensic}.  
Since there are differences in MRI techniques and grading systems for maturity assessment, all studies cannot be compared with each other. Three of the original studies use MRI techniques and grading systems that are more or less comparable to the method used by RMV \cite{kramer2014forensic}, \cite{fan2016forensic}, \cite{ottow2017forensic}. 
However, the relatively small number of participants in relevant ages and shifting results make it hard to regard this method as validated. More and larger studies are needed.

Another aspect of validity is the application of the methods. Validation of assessments is an obvious practice in the field of medicine. Normally, an apprentice makes assessments under the supervision of an experienced assessor. When the rate of correct assessments is sufficiently high, the apprentice is allowed to make them on his or her own. At least for the maturity assessments of MRI knee, RMV has not presented any external validation prior to the large amount of assessments they now have performed. 

In 137 cases where RMV assessed the knee as mature, an external second opinion has been performed by German scientists\footnote{https://www.svd.se/rmv-andrar-aldersbedomning-efter-granskning}. 
These scientists are the ones who have developed and continued to study an MRI knee method close to the one RMV uses. In 75 of these 137 cases (55\%), the German scientists came to the opposite conclusion that the knee was not mature. The cases that has undergone second opinions are the result of private initiatives and they are thus not randomly selected. Therefore one cannot generalize these results to all people who have had their knees assessed as mature. However, one cannot exclude general discrepancies and since RMV uses these German studies as the most important foundation for their method, the results are alarming and require a thorough analysis of the validity of the Swedish assessments. When faced with this criticism RMV performed an analysis of reliability, but no analysis of validity has yet been performed.

We are thus facing a situation where there is substantial uncertainty about the true relationship between 
chronological age and the age indicators used by RMV. There is of course also a large uncertainty about the true
age distribution of the population on which the procedure has been performed. The only firm evidence is 
the information presented in Table~\ref{tab:data}. 
In this paper, we show how simulation within a Bayesian framework may be used to obtain information 
about the possible combinations of population profiles and age indicator models that may explain this data. 
We also show how one may obtain some information about likely classification error rates in such a situation. 

A simple statistical approach to medical age assessment is the following: An age indicator that can take on discrete values 
$I_1,\dots, I_n$ is measured on a study population with known chronological ages. The study population is subdivided according 
to the age indicator, and the chronological ages within each subgroup are modelled with some statistical model, possibly just a normal 
distribution. Then, this statistical model is used to assess the chronological age of persons whose observed age indicator corresponds to the group. 

The main drawback of this simple and common approach is that it assumes that, a priori, the distribution of the ages of the assessed persons 
corresponds to the distribution of ages in the study population. This is clearly not the case in most applications of age assessments, as the assessment is generally 
triggered by circumstances related to age. For example, an immigration authority may decide to require medical age assessment of all asylum seekers, 
of asylum seekers they believe might be re-classified by the assessment, or of asylum seekers whose age they are fully convinced are above the relevant 
age limit of 18 years. Clearly different decisions will lead to different rates of erroneous classification, something that cannot be captured by the 
simple statistical procedure above. 

In this paper, we instead use the following procedure: For each age indicator, we use studies where the indicator has been observed 
in study populations with known chronological ages to establish a statistical model predicting the value of the age indicator as a function of 
chronological age. When assessing the age of a person with an observed age indicator, we combine an a priori distribution for the persons age with the likelihood 
provided from the age indicator and the statistical model to obtain the a posteriori distribution for the age of the person. 
This Bayesian approach is discussed for example in \cite{taroni2010data} (relating to forensics in general) and for example in \cite{thevissen2010human} (relating to age assessments). 

For each of the two age indicators appearing in this paper, we thus need to establish a statistical model predicting the 
value of the age indicator from chronological age. General models are discussed in Section~\ref{secmeth}. 
How to obtain model parameters from publshed studies is discussed in Section~\ref{sec:modelparams}. 
In this paper, we assume that, given chronological age, the probability for observing various values of one indicator
is independent of the value observed for another indicator. The relationship between maturation in the knee and the third molars has not been studied, but the study \cite{gelbrich2015combining} on the wrist and third molars found no correlation 
between these age indicators. Thus, an assumption of conditional independence between the third molars and the knee is not implausible. 

For the approach above to work, one needs to establish an a priori distribution for the age of the person that is assessed. 
In case work, such a distribution will be based on the circumstances of that person, and may vary from case to case. In this paper, we consider data derived from age assessment of 9280 males, and we use a common a priori distribution for these, based simply on the fact that they have in a sense been required\footnote{The tests are not mandatory, and only those who have not been able to make their under-age plausible are offered the tests. However, if they then do not agree to take the tests, they will most likely be assessed as adults.} to submit themselves to 
the medical age assessment procedure arranged by RMV in Sweden. 

In initial computations, we use indicator model parameters derived directly from published studies; 
for MRI knee we used \cite{ottow2017forensic} while for x-ray teeth either 
\cite{lucas2016dental} or \cite{mincer1993abfo}. Combining these with a particular 
fixed population profile results in models that 
do not fit the observed data. Thus it is clear that we need to relax assumptions, either about the indicator models, about the 
age distribution, or both. 

In Section~\ref{sec:modelparams} we set up a series of reasonable 
possible priors for age indicator models, taking the published age indicator studies as starting point. 
In Section~\ref{sec:pop} we discuss population profiles, setting up a hierachical prior designed to minimize the 
effect on results of prior guesses about the profile. In the results in Section~\ref{sec:resultsM1}
we combine the indicator model parameter priors with the hierachical 
population prior and study the properties of the resulting models. 

Any statistical investigation rests on a set of assumptions. In some contexts, it may be easy to agree on reasonable assumptions, the exact nature of which many then be discussed little or not at all. In our context, many reasonable sets of assumptions are possible, with each set leading to somewhat different results. We have confronted this challenge by presenting several possible models in the main text, and also discussing a range of alternatives in the supplementary material 
\cite{supplementary2018} for this paper. Even if results vary somewhat with different models, there are some general conclusions we believe can be drawn, and we present these in Section~\ref{sec:conclusions}. 

\section{Methods}

In Section~\ref{secmeth} we present the stochastic model enabling us to do the computations specified above. In Section~\ref{sec:modelparams} we present the models we use for teeth and knee age indicators while 
Section~\ref{sec:pop} contains a discussion on how we model the age distribution.  
Finally, Section~\ref{sec:technical} contains some technical information surrounding simulation with our model. 

\subsection{Stochastic model}
\label{secmeth}

We assume $K$ different age indicators are observed. We assume age indicator $k$ ($k=1,\dots,K$) can take on $n_k$
different discrete values, denoted $I_{k1},I_{k2},\dots,I_{kn_k}$. 
For each age indicator $k$, we assume there is a model with
parameters $\theta_k$ relating the chronological age $x$ of a person to the
probabilities $p_{kj}(x\mid\theta_k)$ of observing indicator $I_{kj}$, so that we assume
$$
\sum_{j=1}^{k_i}p_{kj}(x\mid\theta_k)=1
$$
for all $x$. 

As an example, assume age indicator $k$ has two different values, $I_{k1}$ representing ``immature'' and $I_{k2}$ representing ``mature''.
In some cases, a reasonable parametric model may be
\begin{equation}
p_{k2}(x\mid\theta_k)= \Phi\left(\frac{x - \theta_{k1}}{\theta_{k2}}\right)
\label{eq0}
\end{equation}
where $\theta_k=(\theta_{k1},\theta_{k2})$ and $\Phi$ is the Probit function (i.e., the cumulative distribution
function for the standard normal distribution). We will consider models where age indicator $k$ can have a third
possible value, $I_{k3}$, representing ``not assessible''. In fact, a model with a constant probability 
for such missing data does not fit the data considered in this paper. Thus, we use instead a linear dependency 
of lack of data on age: 
\begin{eqnarray}
  p_{k3}(x\mid\theta_k) &=& \theta_{k3} + \theta_{k4}(x-20) \label{eq1first} \\
  p_{k2}(x\mid\theta_k) &=& \left(1-p_{k3}(x\mid\theta_k)\right)\Phi\left(\frac{x-\theta_{k1}}{\theta_{k2}}\right)
\label{eq1}
\end{eqnarray}
where now $\theta_k=(\theta_{k1},\theta_{k2},\theta_{k3},\theta_{k4})$. 

For each age indicator $k$ we use a probability density on the space of possible parameters $\theta_k$ to model
the uncertainty in the model. Specifically, consider the model of Equations~\ref{eq1first} and \ref{eq1}. 
As it is reasonable to think that, given age, lack of data is independent of maturity of the indicator, we may write
\begin{equation}
\pi(\theta_k) = \pi(\theta_{k1},\theta_{k2})\pi(\theta_{k3},\theta_{k4}). 
\label{eq2}
\end{equation}
In our setting, the parameters $\theta_{k3}$ and $\theta_{k4}$ concerning lack of data will be 
well informed by the data we are considering, so we will use flat priors $\pi(\theta_{k3},\theta_{k4})\propto1$ for these. 
The priors $\pi(\theta_{k1},\theta_{k2})$ will be based on information obtained from various published studies and will be further 
discussed in Section~\ref{sec:modelparams}. We now define a joint prior 
$$
\pi(\theta) = \pi(\theta_1)\pi(\theta_2)\dots\pi(\theta_K)
$$
where $\theta = (\theta_1,\dots,\theta_K)$. 

As mentioned above, we assume in this paper that, given chronological age $x$, the probability for observing 
various values of one indicator is independent of the value observed for another indicator. Thus, 
assuming a person has age $x$, that a vector $v=(z_1,\dots,z_K)$ of the $K$ different age indicators is 
observed for this person, and given a value for $\theta$, the probability of observing $v$ can be written as 
$$
p(v\mid x,\theta) = \prod_{k=1}^K p_{kz_k}(x\mid\theta_k).
$$

Aside from the parameters of the age indicator observation models used, the major uncertainty in our situation lies in the
distribution of chronological ages in the population on which the observation procedure is applied. In this paper, we will 
use a discretization, using the vector $\{x_1,\dots,x_T\}$ to represent $T$ possible age values. A population profile is then 
represented by a vector $\psi=(\psi_1,\dots,\psi_T)$, with $\psi_i$ indicating the probability for age $x_i$, so that 
$\sum_i\psi_i=1$. We will use a Dirichlet prior on $\psi$, with 
$$
(\psi_1,\dots,\psi_T)\sim\operatorname{Dirichlet}(\alpha/T, \dots,\alpha/T).
$$
for some parameter $\alpha$. Under this prior, the expected value of each $\psi_i$ is $1/T$. Starting with some 
distribution with cumulative density function $F$ which can be considered reasonable, we choose the 
$x_i$ so that $F(x_i) = i/T$. Thus the uneven spread of the $x_i$ will reflect the population profile specified by $F$. The uncertainty around this target distribution is governed by the parameter $\alpha$: We get that 
$$
\psi_1+\dots+\psi_i \sim\operatorname{Beta}\left(\frac{i}{T}\alpha, \frac{T-i}{T}\alpha\right)
$$
so that when $\alpha\to\infty$ we get increasingly little variation around the target distribution, while $\alpha\to0$ gives
increasing flexibility. 

To make computations, we include in the model a variable with information about the actual ages of the persons subjected to the age assessment procedure. Let $v_1,v_2,\dots,v_V$ be the possible values that an age indicator vector $v$ can take 
on, so that $V = n_1n_2\cdots n_K$. Now let $\tau(v_i,x_j)$ represent the count of persons of age $x_j$ having observational vector $v_i$, and let 
$$
\tau = \left\{\tau(v_i,x_j)\right\}_{i = 1,\dots,V;  j=1,\dots,T}
$$
so that $\tau$ is the collection of all these counts. Fixing $\theta$ and $\psi$, $\tau$ has a multinomial distribution, 
$$
\tau\mid\theta,\psi\sim\operatorname{Multinomial}\left(N, \left\{r(v_i,x_j\mid\theta,\psi)\right\}_{
i=1,\dots,V; j=1,\dots,T}\right)
$$
where $N$ is the total number of persons observed and 
$$
r(v_i,x_j\mid\theta,\psi) = \psi_i p(v_i\mid x_j,\theta)
$$
is the probability that a person has age $x_j$ and observational vector $v_i$. 

The actual observations are contained in the vector $y=(y_1,\dots,y_V)$ where, for $i=1,\dots,V$, 
$$
y_i = \sum_{j=1}^T\tau(v_i, x_j).
$$
We have now formulated a full stochastic model for our variables: 
$$
\pi(y,\tau,\theta,\psi) = \pi(y\mid\tau)\pi(\tau\mid\theta,\psi)\pi(\theta)\pi(\psi).
$$
Our strategy is to simulate from this joint distribution conditional on the observed data $y$ using a Metropolis 
Hastings algorithm. There are three different updating steps, where each of the variables $\tau$, $\theta$, and $\psi$ are 
updated while the other variables are kept fixed. 

For $\tau$, we get 
$$
\pi(\tau\mid y, \theta,\psi) \propto \pi(y\mid\tau)\pi(\tau\mid\theta,\psi)
$$
and as $\pi(y\mid\tau)$ simply restricts the sums of counts in $\tau$, we get for $i=1,\dots,V$ that 
$$
\left(\tau(v_i,x_1),\dots,\tau(v_i,x_T)\right)\sim\operatorname{Multinomial}\left(y_i, 
\left\{\frac{r(v_i,x_j\mid\theta,\psi)}{\sum_{k=1}^Tr(v_i,x_k\mid\theta,\psi)}\right\}_{j=1,\dots,T}\right).
$$

For $\theta$ we get 
\begin{eqnarray*}
\pi(\theta\mid y,\tau,\psi) &\propto & \pi(\tau\mid\theta,\psi)\pi(\theta) \\
&\propto & 
\left[\prod_{i=1}^V\prod_{j=1}^Tr(v_i,x_j\mid\theta,\psi)^{\tau(v_i,x_j)}\right]
\prod_{k=1}^K\pi(\theta_k)\\
&\propto&
\left[\prod_{i=1}^V\prod_{j=1}^Tp(v_i\mid x_j,\theta)^{\tau(v_i,x_j)}\right]
\prod_{k=1}^K\pi(\theta_k)
\end{eqnarray*}
which splits as a product over the different age indicators: 
$$
\pi(\theta_k\mid y,\tau,\psi) \propto\pi(\theta_k)\prod_{z_k=1}^{n_k}\prod_{j=1}^Tp_{kz_k}(x_j\mid \theta_k)^{\tau'(z_k, x_j)}
$$
where 
$$
\tau'(z_k,x_j) = \sum_{z_1=1}^{n_1}\dots\sum_{z_{k-1}=1}^{n_{k-1}}\sum_{z_{k+1}=1}^{n_{k+1}}\dots\sum_{z_K=1}^{n_K}
\tau((z_1,\dots,z_K),x_j).
$$
In other words, the posterior probability for a parameter vector $\theta_k$ is proportional to its prior probability times
the product of the probabilities of observing each of the $n_i$ indicator values at each of the possible ages to the power of the count of the persons having this age and indicator value. 

Using a random walk proposal function in the Metropolis Hastings procedure, we can calculate the acceptance probability at each stage. See Section~\ref{sec:technical} for details. 

For $\psi$ we get
$$
\pi(\psi\mid y, \tau,\theta) \propto \pi(\tau\mid\theta,\psi)\pi(\psi) \\
\propto
\pi(\psi)\prod_{j=1}^Tq(x_j\mid\psi)^{\tau''(x_j)} = \pi(\psi)\prod_{j=1}^T\psi_j^{\tau''(x_j)}
$$
where 
$$
\tau''(x_j) = \sum_{v_i=1}^V\tau(v_i,x_j).
$$
Using the Dirichlet prior $\pi(\psi)$ mentioned above, we may simulate $\psi$ from
$$
\psi\mid\tau\sim\operatorname{Dirichlet}\left(\tau''(x_1)+\alpha/T,\dots,\tau''(x_T)+\alpha/T\right).
$$

\subsection{Age indicator model parameter values}
\label{sec:modelparams}

We now turn to how we can obtain estimates $\hat{\theta}_{k1}$ and $\hat{\theta}_{k2}$ for the parameters of 
Equation~\ref{eq0} from published studies on the age indicator. Given the raw data from such a study, 
i.e., a list of pairs of observed chronological ages and age indicators, one may use maximum likelihood 
to fit a model like that of 
Equation~\ref{eq0} and thus obtain an estimate for the model parameters. However, age indicator studies tend to not 
publish their raw data and a more indirect approach is necessary. We have chosen to in each case construct a plausible raw 
data set based on the information in the paper, and then estimate parameters based on this. As ways of obtaining 
such raw data is not the main focus of this paper, we have chosen fairly ad-hoc procedures. 

In \cite{lucas2016dental}, a total of 1000 males are examined, subdivided according to age into 20 groups of 50 males each. 
Each group consists of persons with ages within a specified half-year interval; we approximated the ages to the middle value 
of each such interval. Table 2 in \cite{lucas2016dental} reports the number of males within each group that have 
"mature" teeth. Maturity is defined in terms of stage H of Demirjian's scale, as for the RMV procedure, but using 
only the left mandibular third molar, when available. With no explicit information about the right mandibular third molars in the study, we have chosen to ignore this slight difference in definitions of age indicators. 

With a raw data set reconstructed in this way, 
we can apply maximum likelihood estimation, obtaining the estimates $\hat{\theta}_{11L}=18.6$ and 
$\hat\theta_{12L}=0.7$.  Details of our computations can be found in the available R code. According to Equation~\ref{eq0}, 
we can interpret the result for example as follows: 
We expect mature teeth in about half of those aged around 18.6, about 85\% of  those aged around $18.6 + 0.7 = 19.3$, and about 
97.5\% of those aged around $18.6 + 2\cdot0.7 = 20.0$. 

In \cite{mincer1993abfo}, the male study population has 271 individuals. The number of males observed with each of the 
age indicators D, E, F, G, H for the mandible is not reported, but we interpolate these values from Table 1 in the paper, 
obtaining 37, 43, 45, 55, and 91, respectively. The quantiles of the ages of the persons in each of these five groups are reported in 
Table 3 of the paper. In order to construct actual ages within each group, we construct a picewise linear transformation from the 
quantiles of the standard normal distribution to the quantiles reported in the table. Starting with values evenly spread
according to the standard normal, we apply the transformation to obtain age values, making sure the age limits 
of 14.1 and 24.9 reported in the study are observed. With a raw data set reconstructed, we apply maximum likelihood estimation to obtain the parameters $\hat{\theta}_{11M}=20.0$ and $\hat{\theta}_{12M} = 3.2$. Note that the parameter estimates based on \cite{mincer1993abfo} and \cite{lucas2016dental} are quite different. 

Turning to knees, we derive parameter estimates from \cite{ottow2017forensic}, as this is by far the largest study on MRI knee that uses a 
method comparable to that of RMV. Five different stages of the age indicator occur in the study: IIc, IIIa, IIIb, IIIc, and IV. Table 3 in \cite{ottow2017forensic} lists the number of males in the study population for which each indicator value has been observed, and also gives summary statistics for the ages within each group. We use from these the minimum, maximum, and the three quartile values. With these values as starting point, we reconstruct raw data in a similar way as for \cite{mincer1993abfo}, and apply maximum likelihood estimation. 
The resulting values are $\hat\theta_{21}=18.5$ and $\hat\theta_{22}=1.5$. 
As mentioned in the introduction, a small subset of RMV’s cases has undergone second opinion. In these, more than half of the knees assessed as mature by RMV was deemed immature in the second assessment. One cannot draw general conclusions from such a small and selected subpopulation, but a worst case scenario might be that around half of all knees that are ”almost mature” (i.e. stage IIIc) are incorrectly classified as mature in RMV’s material. Based on that possibility, we have also 
estimated parameters under the assumption that half of those 32 observations classified as stage IIIc are counted together with staget IV as mature. Under this assumption
we get the estimates $\hat\theta_{21c}=17.8$ and $\hat\theta_{22c} = 1.7$. 

In contrast to the relatively few studies that exist on age assessment with MRI knee, there are numerous studies on the third molars. We have not made a comprehensive review of these, and the choice of Mincer and Lucas can probably be discussed as to whether they are the most relevant ones. We have chosen these for our analysis as they are among the few studies on teeth RMV refers to in 
the description of their methodology, \cite{RMV2017}. Thus, these studies have been used to motivate RMV’s choice of methods and ought therefore be relevant to their procedure. 
As results based on the 
Mincer and Lucas studies are quite different, they cannot both correspond to the RMV procedure, but they, and the Ottow study, provide information about what parameter values are reasonable.

Technically, we deal with this 
by defining prior distributions using the values from the studies in different ways. 
For the teeth parameters, we define three priors. The first is centered close to the Lucas parameter estimates and prescribe 
limited variation around these. Mathematically we define 
$$
\pi_{\text{Lucas}}(\theta_{11},\theta_{12}) \propto
\operatorname{Normal}(\theta_{11}; 18.6,   0.2) 
\cdot\operatorname{Normal}(\theta_{12}; 0.7,   0.2)\cdot I(\theta_{12}>0).
$$
In other words, the prior density is proportional to the product of two normal densities, truncated so that $\theta_{12}$ is 
positive. A way to understand this prior is that we are approximately 95\% sure $\theta_{11}$ is in the interval 
$[18.6 - 0.4, 18.6 + 0.4]$ and, independently, approximately 95\% sure $\theta_{12}$ is in the interval $[0.7 - 0.4, 0.7 + 0.4]$. 
We also define a similar "Mincer" prior that is centered on the values $\hat\theta_{11} = 20.0$ and $\hat\theta_{12}=3.2$, and 
a "wide" prior that is centered on values averaged from the two studies, using 4 times as large standard deviation as the 
Lucas and Mincer priors. Parameter values and 95\% credibility intervals are listed in Table~\ref{tab:modeltab}. 

For the knee parameters, we also define three priors. 
The first two are centered close to the Ottow parameter estimates, with the first 
having the same narrow variation as the Lucas and Mincer priors, and the second having the wider variation. 
Finally, we define a third narrow prior centered close to the estimated parameters $\hat\theta_{21c}= 17.79$ and 
$\hat\theta_{22c}=1.76$ obtained by interpreting half of observations in stage IIIc as mature. Parameter 
values and 95\% credibility intervals are listed in Table~\ref{tab:modeltab}. 

\begin{figure}
\begin{center}
\centerline{\psfig{figure = 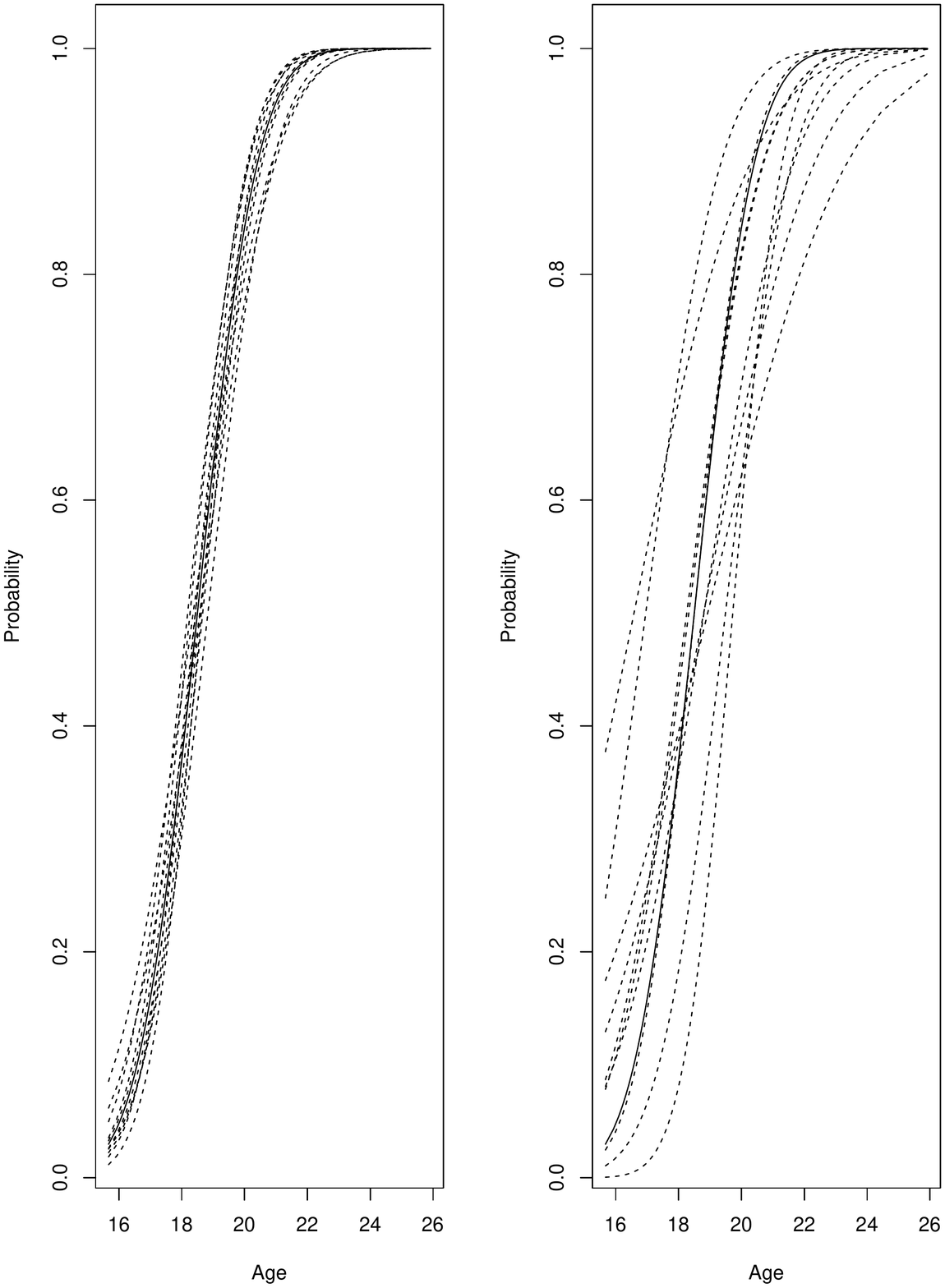, width = 13cm, height=8cm}}
\caption{The probability for observing mature knees as a function of age, assuming the observation is not missing. The continuous line in either plot illustrates the most likely model, with parameters $\hat\theta_{21}=18.5$ and $\hat\theta_{22}=1.5$. The dotted lines illustrate other possibilities under each prior: The left figure uses the narrow Ottow prior while the right figure uses the wide prior.}
\label{fig:priorcompare}
\end{center}
\end{figure}

Figure~\ref{fig:priorcompare} illustrates the narrow and wide knee parameter priors centered on the Ottow estimates by plotting a sample of possible 
age indicator probability curves under each prior. The middle solid curve in each of the plots corresponds to the 
parameters $\hat\theta_{21}=18.5$ and $\hat\theta_{22}=1.5$. The dotted lines indicate how these curves may vary under each 
prior. 

\subsection{Specification of prior for the population profile}
\label{sec:pop}

Specification of a prior age distribution for the population of males that have been subjected to RMVs age determination procedure during 2017 is a difficult task. The prior for an individual should be based on all knowledge about this individual excluding the age indicators. Such general knowledge can include other observations of biological maturity made by medical personnel, observations of psychological maturity made by teachers or other qualified observers, documentable circumstances surrounding the life situation, as well as of course the reasons why the person has been required by the Swedish migration authority to complete the RMV procedure. The age prior for the 
whole population studied in this paper should represent an average over their individual priors.  

The difficulties with establishing such a prior can lead some researchers to the conclusion that frequentist statistical methods where a prior does not seem to be needed are preferable to our Bayesian approach. However, when conclusions are drawn using such 
frequentist methods, they generally correspond to the use of a particular prior, as mentioned in the introduction. For example, 
the "hidden" prior assumption may be that the a priori age distribution corresponds to that of a study population, or that it 
is uniform within some age interval, for example between 14 and 25. So, the relevant question is whether we can establish a prior 
that is more realistic than such hidden priors. 

In this paper, we will use as a starting point an age profile illustrated with the solid line in the upper left-hand plot of Figure~\ref{FinalFigure3}. 
This line represents
the cumulative prior probability that a person has an age below that given on the x-axis. Thus, for example, the prior 
probability that a person is below 18 years is about 35\%. Mathematically, the 
prior is represented by a Gamma density with parameters 4 and 1, translated to be at least 15 and truncated 
to the interval between 15 and 30.  As this distribution is rather 
arbitrarily chosen, we use a hierarchical prior with a lot of uncertainty around this starting point. The exact mathematical 
mechanism for specifying the hyperprior is discussed in Section~\ref{secmeth}. We use here the hyperparameter
$\alpha=3$ together with a discretization into $T=100$ ages. 
As can be seen in the upper left-hand plot of Figure~\ref{FinalFigure3}, 95\% 
credibility intervals for the prior percentage of persons that 
have reached specific ages are quite wide. Thus, we believe the prior is flexible enough so that our choice of starting point 
will not influence results significantly; this is further discussed in the supplementary material \cite{supplementary2018}. 

\subsection{Convergence and accuracy}
\label{sec:technical}

The MCMC simulation outlined in Section~\ref{secmeth} uses a proposal function alternating between changing the population profile $\phi$, changing the tooth model parameters $\theta_1$, and changing the knee model parameters $\theta_2$. Whereas Section~\ref{secmeth}
lists the exact conditional distribution for $\psi$, we use symmetric proposal functions for $\theta_1$ and $\theta_2$, computing also 
acceptance probabilities. Specifically, the proposals perturb the four parameters of each $\theta_k$ using normal distributions, see the 
R code for details. Our choices led to slow but acceptable mixing rates. 

With clearly unimodal posteriors, convergence is fairly easy to assess using plots and multiple chains with different starting points. 
In our implementation, acceptance rates are around  0.1-0.2 for 
both teeth and knee parameters. For final results we simulated 1 million MCMC cycles for each model, using a burn-in of 20000 cycles. 
Each such computation took around 20 minutes on a laptop. However, similar accuracy was obtained with much smaller number of iterations. 
The R code used is available from the supplementary material \cite{supplementary2018}. 

\section{Results}
%\subsection{Results for models M1L and M1M}
\label{sec:resultsM1}
\label{sec:resultsM2}

\begin{figure}
\begin{center}
\centerline{\psfig{figure = 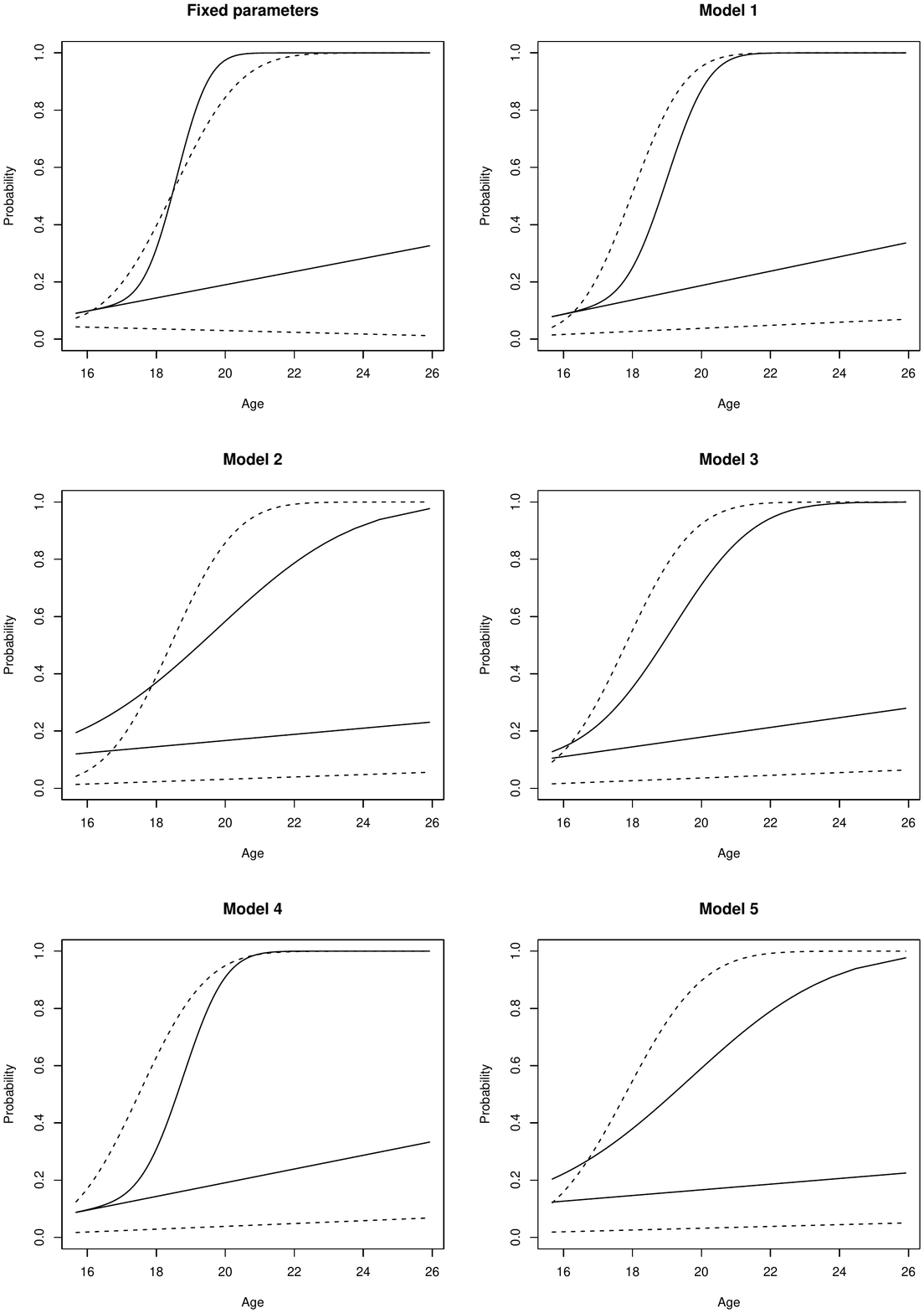, width = 13cm ,height=15cm}}
\caption{Illustration of the posterior expected model parameters, under various models. The straight lines indicate the posterior rates for missing data. The dotted lines correspond to the knee model; the continuous lines to the teeth model.}
\label{FinalFigure2}
\end{center}
\end{figure}

In intial computations, we fix knee parameters to those estimated from the Ottow study (see Section~\ref{sec:modelparams})
and use as a fixed age profile the starting point profile presented in Section~\ref{sec:pop}. 
We then combine with teeth parameters estimated from either the Lucas or the Mincer study 
(see Section~\ref{sec:modelparams}) and in each case estimate the remaining model parameters using maximum likelihood. 
For example, when using the Lucas numbers, we get $\hat\theta_{13} = 0.19$, $\hat\theta_{14} = 0.023$, $\hat\theta_{23}= 0.03$, and $\hat\theta_{24}= -0.003$. Recall that 
$\hat\theta_{13}$ represents the probability for a person aged 20 to have missing tooth data, while $\hat\theta_{14}$ represents the 
slope of the line indicating how this probaility changes with age. $\hat\theta_{23}$ and $\hat\theta_{24}$ are the similar parameters for the knee. 
The parameters are illustrated in the upper left-hand corner of Figure~\ref{FinalFigure2}. 
 
Using either the Lucas or Mincer numbers, we can now make tables of predicted counts, similar to Table~\ref{tab:data}. The predicted tables, however, are
quite different from Table~\ref{tab:data}. Indeed, in both cases, we can reject with p-values smaller than $10^{-5}$ that 
the data in Table~\ref{tab:data} can come from the models, and we can safely discard these initial models. Details of the 
computations can be found in \cite{supplementary2018}.  

The natural question to ask is then: 
Which combinations of age distributions and age indicator models could reasonably have produced the observed data? 
A little thought will convince the reader that there will be no unique solution. For example, if one age profile and set of indicator models could produce the 
data, then simply translating the ages in the age profile and the models by the same amount will give a result that fits the 
data equally well. So our question needs to be refined further, we need to ask: Which combinations of reasonable
age distributions and reasonable age indicator models could reasonably have produced the observed data? 

It is possible to make computations where we relax only assumptions about the population profile or only those about the age indicator model parameters. 
Such computations can be found in \cite{supplementary2018}. However, the most realistic option seems to be to relax assumptons about both 
model components simultaneously. The population profile is certainly unknown and will probably not look exactly as the starting point profile of 
Section~\ref{sec:pop}. So from now on, we use the hierachical prior of Section~\ref{sec:pop}. The age indicator model 
parameters are also uncertain; for example, the Mincer and the Lucas parameter estimates cannot both be correct for the RMV procedure. However, there may 
be different opinions about the information the studies contain about what the reasonable RMV parameters are.
Thus, we have chosen to report results for five different models, each using the information from the age indicator studies in different ways.  

\begin{enumerate}
\item[Model 1] combines the prior centered on the Lucas parameter estimates with the prior centered on the Ottow parameter estimates. 
\item[Model 2] is the same as model 1 except that the Lucas numbers are replaced with the Mincer numbers. 
\item[Model 3] recognizes the large uncertainty in both knee and teeth parameters and uses the wide prior centered on the average between the Mincer and Lucas parameters together with 
the wide prior centered on the Ottow parameter estimates.  
\item[Model 4] combindes the prior centered on the Lucas parameter estimates with the prior centered on the adjusted Ottow parameters computed in Section~\ref{sec:modelparams}. 
\item[Model 5] is the same as model 4 except that the Lucas numbers are replace with the Mincer numbers. 
\end{enumerate}

\begin{table}
\begin{small}
\begin{center}
\begin{tabular}{|l|c|c|c|c|c|c|c|} \hline
          &                            &  Prior                 & Posterior             & Posterior            & Posterior       &    Posterior  & Posterior  \\
          &                            &                          &    Model 1              &  Model 2                  & Model 3               &     Model 4        & Model 5        \\ \hline
Lucas  & $\theta_{11}$       &  18.6                  &     19.1              &                          &                     &    18.8         &              \\
          &                            & 18.2 -- 19.0       &  18.8 -- 19.3    &                          &                           &  18.5 -- 19.1  &            \\    
          & $\theta_{12}$      &  0.7                   &       0.9              &                           &                      &      1.0          &               \\
          &                            &  0.3 -- 1.1        &  0.8 -- 1.2       &                           &                           &   0.8 -- 1.2    &                \\ \hline
Mincer & $\theta_{11}$      &     20.0              &                          &    20.0               &                      &                       &    19.9     \\
          &                            &   19.6 -- 20.4    &                         &  19.7 -- 20.3     &                      &                      & 19.6 - 20.3  \\    
          & $\theta_{12}$      &       3.2              &                         &       3.1               &                        &                     &   3.2           \\
          &                            &   2.8 -- 3.6       &                         &   2.8 -- 3.5       &                        &                      & 2.8 -- 3.6  \\ \hline
Wide   & $\theta_{11}$      &  19.3                 &                          &                         &        19.3          &                        &                   \\
prior   &                            & 17.7 -- 20.9      &                          &                          & 18.3 -- 20.4   &                        &                      \\    
teeth   & $\theta_{12}$      &        2.0              &                         &                          &        1.9           &                        &                      \\
          &                            &  0.4  --  3.6      &                          &                          &     0.9 -- 3.2   &                        &                      \\ \hline\hline
Ottow & $\theta_{21}$      &     18.5              &   18.0                 &         18.5           &                         &                       &                      \\
          &                            &   18.1 -- 18.9     &   17.7 -- 18.3    &    18.1 -- 18.8  &                        &                        &                      \\    
          & $\theta_{22}$      &       1.5              &   1.2                   &        1.5             &                         &                        &                       \\
          &                            &   1.1 -- 1.9       &   1.0 -- 1.6       &    1.1 -- 1.8      &                         &                      &                      \\ \hline
Wide   & $\theta_{21}$      &  18.5                 &                          &                          &       17.9             &                       &                       \\
prior   &                            & 16.9 -- 20.1       &                           &                          &    16.9 -- 18.9  &                        &                       \\    
knees  & $\theta_{22}$      &        1.5             &                          &                          &        1.5              &                         &                       \\
          &                            &  0.0  --  3.1      &                         &                           &   1.0 -- 2.5       &                          &                        \\ \hline
Ottow  & $\theta_{21}$      &  17.8                 &                          &                        &                            &       17.5            & 17.8                \\
IIIc     &                            & 17.4 -- 18.2     &                           &                        &                          &    17.2 -- 17.8   & 17.5 -- 18.2  \\    
          & $\theta_{22}$      &        1.7             &                          &                         &                           &        1.5              & 1.7                \\
          &                            &  1.3 --  2.1      &                         &                          &                           &   1.2 -- 1.9       &  1.4 -- 2.1      \\ \hline
\end{tabular}
\caption{Prior and posterior parameter distributions. The ranges indicate approximate 95\% credibility intervals for 
each parameter.}
\label{tab:modeltab}
\end{center}
\end{small}
\end{table}

The actual age indicator parameter values used in the priors of each model can be read from Table~\ref{tab:modeltab}, 
which also lists information about the posteriors for the parameters. 
A first impression is that the distribution of parameter values do not change very much from the prior to the posterior. 
The clearest change is in model 1, where $\theta_{11}$ and $\theta_{21}$ are very close in the prior and more than one year apart in the posterior. 
In fact, in all the posterior models, the difference between the two parameters is at least 1.1 years, meaning that knees mature on average at least one year 
before teeth. 

\begin{figure}
\begin{center}
\centerline{\psfig{figure = 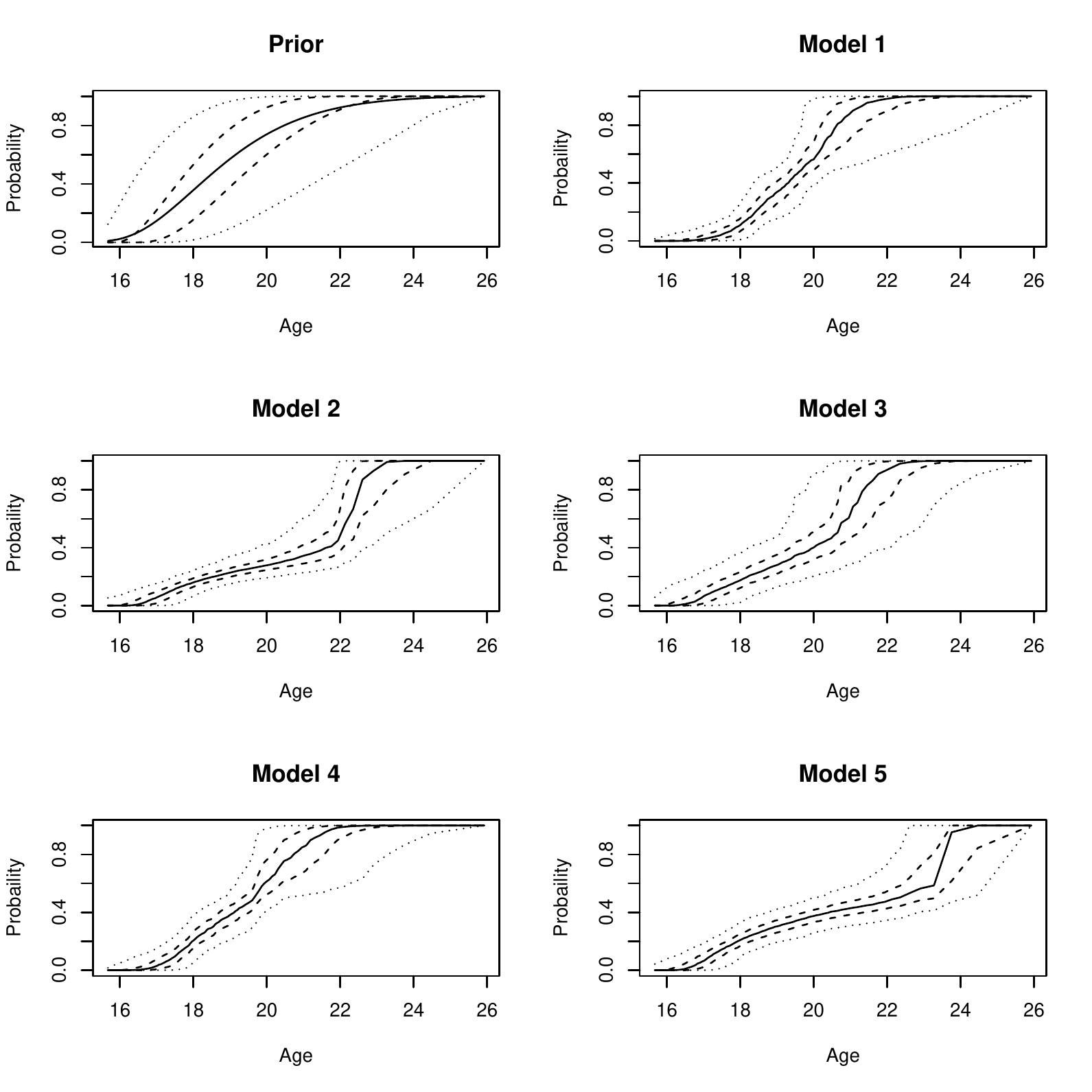, width = 13cm ,height=15cm}}
\caption{Prior and posterior population profiles. The middle line in each plot indicates the most likely profiles. The other lines 
delineate the 2.5\%, 25\%, 75\%, and 97.5\% quantiles, respectively. Thus, vertical intervals between the two dotted lines represent 95\% credibility intervals.}
\label{FinalFigure3}
\end{center}
\end{figure}

To better interpret the models, we also need to look at their posterior age population profiles, illustrated in Figure~\ref{FinalFigure3}. 
As an example, consider the posterior for model 2, which has a curious step-like shape. 
The central black line roughly goes from around 0.4 to around 0.9 over the year between the ages of 22 and 23. 
This means that around half of the population is predicted to have an age between 22 and 23 years. 
In other words, in order to explain the data of Table~\ref{tab:data} using the parameters of model 2, we need to accept that 
roughly half of those going through the RMV procedure were exactly between 22 and 23 years, while there were much fewer 
in any of the other age groups between 15 and 30. In our opinion, this is unrealistic, and would tend to discount our trust in 
model 2. A similar thing happens with model 5, and even to some extent with model 3. Thus, in terms of posterior age profiles, 
we conclude that models 1 and 4 are most realistic, whereas we put less trust in results for model 2 and 5, and to some degree model 3.

How can the population profile effect above be explained? The data of 
Table~\ref{tab:data} show that there are substantially more persons with mature knees and immature teeth than with immature knees and 
mature teeth. For this to happen, there needs to be a substantial age interval where it is substantially more probable that knees are mature than that teeth are mature. 
The illustration of the fixed parameters model in Figure~\ref{FinalFigure2} shows how that model fails to fulfill this criterion. 
The other models fulfill the criterium in different ways. However, we can notice that in models 2 and 5, the line for knee maturity crosses below the 
line for tooth maturity at later ages than in the remaining models. Thus, in models 2 and 5, the ages of those tested need to be more concentrated in a particular short interval where the probability for knee maturity is substantially higher than the probability for tooth maturity. This results in the population 
"bumps" for these models observed in Figure~\ref{FinalFigure3}. 

It also explains why the $\theta_{21}$ parameter generally needs to be 1-2 years smaller 
than the $\theta_{11}$ parameter for models to fit the data. In fact, in models 1 through 4, the posterior difference between these parameters is on average 1.1 to 1.5 years. 
The difference is 2.1 in model 5, but as mentioned above, we put less trust in the results of this model. Further computations, see \cite{supplementary2018}, confirm that 
the time where 50\% of persons have attained a mature knee occurs 1 - 1.5 years before the time when 50\% of persons have attained mature teeth. Specifically, 
this time point for knees cannot occur after this time point for teeth.

\subsection{Classification error rates}
\label{sec:errorrates}

Each of the different models have consequences for the actual error rates of the current RMV procedure.
In each stage of our simulation we can compute the probability for a person in each of the 9 classification groups to be over 18.  
The results are given in Table~\ref{tab:errorrates} in the form of classification error rates, using the classification rule employed by RMV and the Swedish migratory authority. 

\begin{table}
\begin{center}
\begin{tabular}{|l|c||c|c|c|c|c|} \hline
RMV      & N            &      Model        & Model      &       Model       &     Model     &     Model    \\ 
classification &         &           1         &       2        &           3       &           4     &         5      \\ \hline
\multicolumn{7}{|c|}{Error rates when classifying as above 18}                                  \\ \hline
K+, T+   &   4176    &     1         &     1           &          2       &         2      &    3         \\ 
              &               &  0--3       &   0--3      &     0-8           &    0--8      &   0--7       \\ \hline
K+, T-   &   1735     &     15        &     11          &      23         &       38     & 26            \\ 
              &               &   1--46   &    2--27    &     1-68          &    5--80      &  7--54    \\ \hline
K+, T0   &   1364     &    3           &     3            &      6         &          8     &   7        \\ 
              &               &   0-10   &     0--7          &      0-18       &    1--20         &  1--15        \\ \hline
K-, T+   &   348        &   9         &     50           &     35        &         20       &    57      \\ 
              &               &   0--35   &    15--85     &    0--88     &    1--63          &   19--90     \\ \hline
K0, T+   &   187      &    1            &   2               &   3           &           2        &      4      \\ 
              &               &   0--4    &     0--7       &    0--10       &      0--9          &    1--10      \\ \hline
\multicolumn{7}{|c|}{Error rates when classifying as below 18} \\ \hline
K-, T-   &   1087    &    49            &      26           &   28          &      28           &      23     \\ 
              &               &  10--96    &     4 -- 68      &    2--89    &    4--79           &    3--61     \\ \hline
K-, T0   &   237       &   66           &     36             &  43         &         48        &     31    \\ 
              &               &  32--98  &    8--75     &    6 --94   &     17--88           &    5--69    \\ \hline
K0, T-   &   83         &    77           &   78             &        68     &         55          &  64    \\ 
              &               &   45--99  &    57--95   &  27--98       &   17--92          &  36--89     \\ \hline
\end{tabular}
         \caption{Estimated error rates in percent when classifying as over 18. The ranges contain a 95\% credibility interval.}
\label{tab:errorrates}
\end{center}
\end{table}

A first observation may be that error rates are rather large. Although error rates when classifying adults as children seem to be 
larger than those when classifying children as adults, the latter may be considered a larger problem, because of the 
large consequences it has for a child to be classified as an adult. Another observation is the huge uncertainty surrounding 
many error rates. This uncertainty is a consequence of the lack of knowledge about the fundamental parameters of this 
unvalidated age assessment procedure. 

When interpreting these numbers, one should remember how the different models are meant to reflect 
different assumptions about relevancy of different published studies for the RMV procedure. For example, in models 1 and 4, 
one assumes the RMV procedure corresponds more closely to the one employed in Lucas, while in models 2 and 5, one assumes 
it corresponds more closely the the one employed in Mincer. 
Interpretation should also take into account that models 2 and 5 seem unlikely in our study because they 
appear to require a peculiar age profile to fit the data. 

\subsection{Conclusions}
\label{sec:conclusions}

Firstly, it seems clear that the properties of the unvalidated RMV procedure are quite uncertain. Studies that have been used to 
argue that the procedure has certain properties seem to give quite different parameter estimates, and the uncertainties in 
important properties such as the classification error rates are large. In our view, it is unacceptable 
that a procedure with such unclear properties has the central role that RMVs procedure has in Sweden's age assessment of of asylum seekers
and in some criminal cases.  

Secondly, even with these uncertainties, some information can be obtained from the study. For example, it seems a consistent conclusion
is that the age where 50\% of knees are mature occur around 1-1.5 years before corresponding age for teeth teeth, as measured by the RMV procedure, as other models would require an
unrealistic age profile to explain the data. 

Thirdly, even if there are large uncertainties, some information about the error rates can be estimated. For some groups, in particular 
those that have measured one mature age indicator and another immature, these rates seem to be around 10-30\%, and may be 
above 50\%. In our view, 
these rates are unacceptably high, taking into consideration the serious consequences for a child to be
classified as an adult. 

\section{Discussion}
\label{sec:discussion}

In this paper, we present a general method for studying the properties of an unvalidated age assessment procedure. Although 
uncertainties in many results are large, we show how it is possible to obtain some information about model parameters and error rates using only 
classification counts and priors guided by published studies. A key reason % it is possible to obtain results 
is that two age indicators have been observed, so that one can learn 
about the differences in their parameters. Applying our model in a situation with three or more age indicators would probably further strengthen results. 

As with any elaborate stochastic model, one needs to ask how much results depend on the particular technical assumptions and choices we have made. 
For example, we have assumed conditional independence between age indicators given age. Should evidence appear showing that knee and tooth maturity are 
strongly correlated given age, conclusions from this paper would certainly change to some degree. 
The particular form of our age indicator probability model given in Equations~\ref{eq1first} and \ref{eq1} is of course also a simplification. In particular, we do not attempt to adjust for the fact that 
the asylum seekers the RMV procedure has been applied to generally have different genetic and socioeconomic backgrounds compared to the study populations of most age indicator studies.  
Further research on age indicators is likely to yield more complex models better adapted to reality. 
But in the context of the current paper, where model 
uncertainty is already quite high, resulting adjustmens are likely to be moderate. 

We have made a substantial effort to determine the influence on results of our choice for hierarchical prior for the population age profile. 
Although the parametric model is fairly arbitrary, its properties, as illustrated in the top left-hand plot of Figure~\ref{FinalFigure3}, seem reasonable. 
In particular, there is a wide prior uncertainty about the proportion of those tested that has an age below 18. 
Figure~\ref{FinalFigure3} illustrates how the posterior distributions are fairly different from the prior, indicating that the prior has a limited influence. 
The supplementary material \cite{supplementary2018} contains further discussions and studies of the population age profile prior, including results using an alternative prior. 
Based on our work, we are confident that we would reach the conclusions listed in Section~\ref{sec:conclusions} using any reasonable choice of population prior. 

We have in this paper presented numerical results using a handful of different models. Although we believe these span a set of reasonable combinations 
of age indicator model assumptions, there are cleary many more combinations for which one could do simulations. Indeed, we hope that 
the R code developed for this paper can be used for further investigations of the relationship between such model assumptions and resulting error rate 
estimates. The R code is available as part of the online supplementary material \cite{supplementary2018}. 

As the RMV procedure is still in use, deciding the fate of hundreds of people every month, there is an urgent need to learn more about its properties. 
One possibility is to use "second opinion data", now available for some of those 9280 investigated, to calibrate our model. An even more powerful 
calibration could of course be obtained using data where the current RMV procedure is performed on as study population with known ages. 
% Such a validation study has long been promised by RMV, but it has not been published so far. 

Another interesting possibility is to make our modelling of the RMV procedure more detailed. In particular, as RMV uses two experts for each 
age indicator and both experts need to agree on maturity for declaring the measured body part as mature, there is a possibility that 
results depend on selection of observers. However, requests for data from RMV which could could be used in such an extended model 
have so far not been accommodated. 

%SWITCH for IJLM: 
\bibliographystyle{plainnat} 

\bibliography{PaperBib}

\end{document}